\documentclass[conference]{IEEEtran}
\IEEEoverridecommandlockouts
\usepackage{cite}
\usepackage{amsmath,amssymb,amsfonts}
\usepackage{algorithmic}
\usepackage{algorithm}
\usepackage{graphicx}
\usepackage{textcomp}
\usepackage{booktabs}
\usepackage[table,xcdraw]{xcolor}
\usepackage[left=1.25cm,right=1.25cm,top=0.95cm,bottom=1.15cm]{geometry}
\begin{document}
\title{Promise of Data-Driven Modeling and Decision Support for Precision Oncology and Theranostics
{\footnotesize \textsuperscript{}}
}
\author{\IEEEauthorblockN{Binesh Sadanandan$^1$}
\IEEEauthorblockA{\textit{Department of Electrical and Computer}\\ \textit{Engineering and Computer Science} \\
\textit{University of New Haven}\\
West Haven, CT, USA}\\
\and
\IEEEauthorblockN{Vahid Behzadan$^2$}
\IEEEauthorblockA{\textit{Department of Electrical and Computer}\\ \textit{Engineering and Computer Science} \\
\textit{University of New Haven}\\
West Haven, CT, USA}
\thanks{bsada1@unh.newhaven.edu$^{1}$, vbehzadan@newhaven.edu$^{2}$}
} 
\maketitle
\pagestyle{plain}
\IEEEaftertitletext{\vspace{-2\baselineskip}}

\begin{abstract}
Cancer remains a leading cause of death worldwide, necessitating personalized treatment approaches to improve outcomes. Theranostics, combining molecular-level imaging with targeted therapy, offers potential for precision oncology but requires optimized, patient-specific care plans. This paper investigates state-of-the-art data-driven decision support applications with a reinforcement learning focus in precision oncology. We review current applications, training environments, state-space representation, performance evaluation criteria, and measurement of risk and reward, highlighting key challenges. We propose a framework integrating data-driven modeling with reinforcement learning-based decision support to optimize radiopharmaceutical therapy dosing, addressing identified challenges and setting directions for future research. The framework leverages Neural Ordinary Differential Equations and Physics-Informed Neural Networks to enhance Physiologically Based Pharmacokinetic models while applying reinforcement learning algorithms to iteratively refine treatment policies based on patient-specific data.
\end{abstract}

\section{Introduction}

Cancer, as the second leading cause of death worldwide, represents a critical threat to global public health\cite{bray2018global}. Even in highly developed nations like the United States, its increasing incidence underscores the urgent need for continued research and preventative measures\cite{bray2018global}. Advances in precision medicine, particularly within precision oncology, offer innovative strategies to target cancer with greater specificity and improve patient outcomes \cite{jameson2015precision}.

Precision medicine tailors therapeutic strategies to individual patients by accounting for genetic, cellular, and environmental factors. This personalized approach optimizes benefits while reducing costs and complications. In recent years, Radio Pharmaceutical Therapy (RPT) has emerged as a compelling treatment method for various cancer types, delivering radiation systemically or locally through radionuclides and enabling non-invasive imaging of the therapeutic agent's biodistribution.

Theranostics, combining therapy and diagnostics, integrates diagnostic tools with therapeutic agents for simultaneous or sequential visualization and treatment. This approach capitalizes on disease-specific biomarkers coupled with radioactive compounds visible through molecular imaging techniques. A key advantage is the capacity for quantitative pharmacokinetic measurement of molecular drug target uptake, facilitating imaging of therapeutic drug delivery to tumors and normal tissue.

Despite promising outcomes, successful implementation of precision theranostics faces various challenges, including high costs, limited reimbursement, and hesitant attitudes among clinicians\cite{beyer20232022}. Standardized dosing regimens neglect diversity in individual physiological responses, potentially leading to overtreatment or inadequate intervention\cite{emmett2023patient}. There is growing advocacy for personalized dosing strategies using early-response biomarkers to tailor treatment more precisely\cite{emmett2023patient}.

Computational oncology links data-centric analysis with cancer research, utilizing various mathematical and computational methodologies to explore, model, and forecast cancer dynamics. Data-driven modeling extracts knowledge directly from observed data, using advanced algorithms to identify patterns and causative links in complex datasets \cite{zhu2020application, munir2019cancer}.

Reinforcement Learning (RL) has shown promise in automating intricate decision-making for personalized treatment plans. While data-driven methodologies excel at extracting patterns from static datasets, RL extends this by introducing interactive learning and adaptation through trial and error in simulated environments \cite{zhu2020application, munir2019cancer, lapan2018deep, sallab2017deep}.

The contributions of this paper include:
\begin{itemize}
    \item Proposing a framework merging advanced RL with data-driven modeling to create dynamic, personalized oncology treatment plans
    \item Introducing the application of RL algorithms to iteratively learn and refine treatment policies based on real-time patient data and simulated scenarios
\end{itemize}

\section{Radio Pharmaceutical Therapy (RPT)}

A radiopharmaceutical combines a pharmaceutical component with a radioactive isotope, targeting specific biological processes. The drug's chemical structure dictates its biological properties, ensuring it binds to specific cells associated with disease, while the radionuclide provides imaging or therapeutic capabilities. RPT targets and destroys cancerous cells through radiation-induced damage.

In theranostics, $X$-rays and $\gamma$-rays serve pivotal roles in the diagnostic phase, allowing precise mapping of radiopharmaceutical distribution. The therapeutic efficacy hinges on electron emissions (Auger electrons, $\beta$-particles, and monoenergetic electrons) delivering localized, cytotoxic radiation to cancer cells.

Patient screening for theranostic applications uses biochemically identical isotopes—one for diagnostic imaging through PET/CT, another for therapeutic purposes. For example, I-123/I-124 for imaging and I-131 for therapy in thyroid cancer; Gallium-68 for imaging and Lutetium-177 for treatment in neuroendocrine tumors.

\subsection{Current RPT Treatments and Dosing Practices}

FDA-approved clinical agents include Iodine-131 for thyroid conditions, Lutetium-177 for neuroendocrine and prostate cancers, Radium-223 for bone metastases in prostate cancer, Samarium-153 for bone pain palliation, and Yttrium-90 for liver tumors and non-Hodgkin lymphoma. These agents leverage unique expression of cellular receptors or physiological differences in tissue uptake to selectively deliver radiation.

The mechanism of action primarily involves binding to specific cellular receptors overexpressed in cancerous tissues. For example, Lutetium-177 labeled DOTATATE targets somatostatin receptors in neuroendocrine tumors, while PSMA-targeted therapies like Lutetium-177 PSMA target prostate-specific membrane antigen in prostate cancer cells. Radium-223, emulating calcium, selectively accumulates in bone metastases.

Dosing varies widely: Iodine-131 has varying doses for thyroid conditions, Lutetium-177 uses standardized regimens for neuroendocrine tumors, and Radium-223 employs weight-based dosing for prostate cancer with bone metastases.

\subsection{Personalized Dosimetry}

Dosimetry, the methodical quantification of absorbed radiation dose, leverages diagnostic imaging to tailor subsequent treatment cycles. By assessing pharmacokinetics at temporal intervals, clinicians can adjust dosage to optimize therapeutic benefits while mitigating risks to non-targeted tissues\cite{sgouros2020radiopharmaceutical}.

RPT utilizes the absorbed dose (D), representing energy imparted per unit mass of tissue, to dictate therapy's biological impact. This measure is a robust indicator of expected biological response, allowing real-time pharmacodynamic evaluations\cite{sgouros2020radiopharmaceutical}.

Mathematically, dose is defined as $D = \Delta E / \Delta m$, where $D$ is dose in Grays, $\Delta E$ is energy absorbed, and $\Delta m$ is mass of irradiated matter. For RPT, this is refined to $D(T \leftarrow S) = \tilde{A}(S) \times S(T \leftarrow S)$, where $D$ signifies dose absorbed by target organ from radiating source organ, $\tilde{A}(S)$ represents time-integrated activity, and $S(T \leftarrow S)$ is the S-factor calculating dose imparted to the target for each radionuclide decay in the source organ \cite{ramirez2019radiopharmaceutical}.

Expanded, the formula becomes:
\begin{equation}
    D(T \leftarrow S) = \tilde{A}(S) \times \frac{a \times \phi(T \leftarrow S)}{M(T)} 
\end{equation}

where $a$ represents total energy emitted from source organ, $\varphi(T \leftarrow S)$ is fraction of energy absorbed in target organ, and $M(T)$ is mass of target organ \cite{ramirez2019radiopharmaceutical}.

Dosimetry in RPT is complex due to variety of particle types and their distinct interactions\cite{sgouros2020radiopharmaceutical}. Accurate dosimetry requires imaging resolution matching emitted particles' spatial scale of energy deposition. Voxel-level dosimetry is generally adequate for beta emitters but may need more precise measurements for alpha emitters with short range and high ionization density.

\subsection{Decision Support in RPT Dosimetry}

Several commercially available software tools support RPT dosimetry, including QDOSE, PLANET Dose, GE Dosimetry Toolkit, Rapid, and Voximetry Torch\cite{capala2021dosimetry}. These employ three primary approaches:

\begin{itemize}
    \item \textbf{Dose Factor (S value)–Based Calculation}: Uses predefined S-factors representing mean dose absorbed by target organ per unit of activity in source organ, commonly applied in organ-level dosimetry\cite{capala2021dosimetry}.
    \item \textbf{Dose-Point Kernel Convolution}: Convolves dose-point kernels with activity distribution to calculate dose distribution, beneficial for voxel-level dosimetry\cite{capala2021dosimetry}.
    \item \textbf{Monte Carlo Radiation Transport Simulation}: Simulates physical processes of radiation as it travels through matter, accounting for complex geometries and stochastic nature of radiation interactions\cite{capala2021dosimetry}.
\end{itemize}

\subsection{PBPK Approaches}

Physiologically based pharmacokinetic (PBPK) modeling has emerged as a powerful tool particularly well-suited for RPT challenges. PBPK models predict drug efficacy, understand ADMET properties, and inform decision-making throughout drug development\cite{siebinga2021physiologically,gospavic2016physiologically}. Unlike traditional pharmacokinetic models, PBPK approaches construct a detailed 'virtual patient' by representing organs and tissues as compartments with specific physiological characteristics.

To address PBPK models' complexity, the "reaction graph" notation was introduced\cite{fele2024physiologically}, drastically simplifying modeling for radiopharmaceuticals. This approach has been applied to explore competition between radiolabeled and unlabeled ligands, evaluate multi-bolus injection benefits, and analyze how albumin-binding affinities influence targeted dose delivery\cite{fele2024physiologically}.

PBPK models' ability to integrate patient-specific data makes them ideal for guiding personalized RPT treatments\cite{jimenez2021effect,kletting2016optimized}. Predictions of radiation exposure help clinicians tailor dosage to maximize tumor targeting while minimizing adverse effects. Furthermore, these models can evaluate tumor response to therapy\cite{kletting2019modeling}, tracking treatment progress and informing clinical decisions.

The concept of Theranostic Digital Twin (TDT) integrates detailed population-based medical knowledge with patient-specific data to tailor individualized therapeutic strategies\cite{rahmim2022theranostic}. By employing PBPK modeling and computational models, TDT enhances simulation, personalization, and optimization of treatment scenarios, continuously adapting to new patient data to refine therapy decisions\cite{rahmim2022theranostic}.

\section{Open Problems and Challenges}

Personalized dosimetry remains crucial in RPT, particularly for protecting vulnerable organs during therapy\cite{dash2015peptide}. Accurate input data and reliable radiobiological models are essential for improving dosimetry calculations\cite{sgouros2014dosimetry}. While dosimetry-guided RPT demonstrates improved efficacy\cite{thaiss2023personalized}, routine clinical implementation poses challenges.

The theranostics field seeks to integrate targeted diagnostics and therapeutics, requiring advanced algorithms, innovative study designs, and re-evaluation of regulatory processes\cite{mizuno2022future}\cite{goetz2018personalized}. A reliable supply chain of radionuclides is critical for supporting research, clinical trials, and patient access\cite{updegraff2013nuclear}.

Cost-effectiveness considerations are vital for RPT, which often involves high economic and resource costs\cite{thaiss2023personalized}. Identifying individuals most likely to benefit from RPT can significantly enhance the value and efficiency of care provided.

Beyond dosimetry and development, challenges include:
\begin{itemize}
    \item Need for faster preclinical dosimetry tools\cite{gupta2018voxel}
    \item Optimal integration of radiometal isotopes into theranostic designs\cite{notni2018re}
    \item Exploring computational approaches for predicting treatment response\cite{stella2023smart}
    \item Addressing limitations in lesion quantification
    \item Overcoming the inefficacy of one-size-fits-all therapies\cite{spear2001clinical}
    \item Managing limitations of Monte Carlo simulations\cite{reynaert2007monte}
    \item Moving beyond population-level studies for radiopharmaceutical dosing\cite{stokke2017dosimetry}
\end{itemize}

To fully harness RPT's promise, efficient, accurate, and personalized dosimetry methods are urgently needed. Furthermore, the inherent heterogeneity of tumors and their stochastic progression call for adaptable treatment strategies rooted in real-time data and decision optimization frameworks.

\section{Proposal: Data-Driven Modeling for Theranostics}

Our research proposes enhancing PBPK models by integrating Neural Ordinary Differential Equations (Neural ODEs)\cite{chou2023machine}. This approach leverages neural networks' capacity to learn complex, nonlinear dynamics of drug distribution, capturing time-dependent behavior of pharmacokinetic processes accurately.

We discretize the temporal domain into points serving as inputs for Neural ODEs. Parameters within the neural network are initialized stochastically, setting the stage for iterative training that aligns model predictions with physiological responses. The process refines neural network parameters to minimize a loss function encapsulating PBPK system dynamics and initial conditions accuracy.

\begin{algorithm}[!h]
\caption{Algorithm for Optimizing Neural Network Parameters in PBPK Modeling}
\begin{algorithmic}[1]
\REQUIRE Define $t_{\text{batch}}$ as array of discretized time points
\REQUIRE Initialize neural network parameters $\theta$ randomly
\REQUIRE Set initial loss to high value and define tolerance
\WHILE{Loss > tolerance}
    \STATE Solve for steady-state conditions using $g(N_{\text{nss}}, N_{\text{ss}}) = 0$
    \STATE Compute $N_{\text{nss}}$ using neural network with parameters $\theta$
    \STATE Update Loss based on ODE system and initial condition discrepancy
    \STATE Optimize $\theta$ to minimize Loss
\ENDWHILE
\ENSURE Optimized parameters $\theta^*$ solve the PBPK model
\end{algorithmic}
\end{algorithm}

To accurately simulate complex pharmacokinetics, we enhance PBPK models with Neural ODEs. Our methodology models drug interactions across physiological compartments—blood plasma, liver, and kidneys—governed by ODEs capturing mass transport, binding, and metabolism:

\begin{align}
\frac{dP}{dt} &= - k_{p \to l} P + k_{l \to p} L - k_{p \to k} P + k_{k \to p} K \\
\frac{dL}{dt} &= k_{p \to l} P - k_{l \to p} L - k_{\text{met}} L \\
\frac{dK}{dt} &= k_{p \to k} P - k_{k \to p} K - k_{\text{ex}} K
\end{align}

where $P$, $L$, and $K$ represent drug concentrations in plasma, liver, and kidneys, respectively, and the constants represent rate parameters for transport, binding, metabolic, and excretion processes.

We define a vector of drug concentrations $\mathbf{C}_{NN}(t)$ predicted by the neural network at time $t$, encompassing all compartments. The neural network learns dynamics captured by function $\mathbf{f}(t, \mathbf{C}_{NN}(t), \boldsymbol{\theta})$, with $\boldsymbol{\theta}$ denoting physiological parameters.

The network is trained to minimize a loss function combining discrepancies between predictions and true PBPK model dynamics, along with physics-based constraints:

\begin{align}
L_{\text{ODE}} &= \frac{1}{N} \sum_{i=1}^{N} \left\| \frac{d \mathbf{C}_{NN}(t_i)}{dt} - \mathbf{f}(t_i, \mathbf{C}_{NN}(t_i), \boldsymbol{\theta}) \right\|^2 \\
L_{\text{phys}} &= \left\| \sum_{i} C_{NN, i}(t)V_i - D \right\|^2
\end{align}

where $V_i$ denotes volume of compartment $i$, and $D$ represents total administered drug dose.

Our research will use an innovative PBPK simulator developed by \cite{fele2024physiologically} to create high-fidelity datasets for training and evaluating our Neural ODE framework. This simulator offers several advantages:
\begin{itemize}
    \item Controlled parameter variation to explore model sensitivity
    \item Generation of diverse physiological states, dosages, and demographics
    \item Ground truth benchmarking for validating model performance
\end{itemize}

\section{Proposal: Decision Support Systems}

In RPT settings, finding an optimal therapy profile can be modeled as a sequential decision-making problem framed as a Markov Decision Process (MDP). This framework extends the Markov model such that state transitions occur due to agent actions, with each transition yielding a numerical reward as a local measure of usefulness.

We propose developing a Reinforcement Learning (RL)-based Decision Support System (DSS) tailored for Theranostics in RPT\cite{boominathan2020treatment}. Leveraging data-driven personalized PBPK and post-therapy PET scans, this system aims to optimize drug dosing strategies by continually adapting to patient-specific responses.

In this MDP formulation, each state encapsulates the patient's current medical condition—comprising variables from diagnostic scans, treatment history, and data-driven PBPK model. Actions correspond to various drug dosing regimens, allowing real-time adjustments. The RL agent learns an optimal policy mapping states to actions that maximize a reward function balancing tumor reduction with minimizing radiation exposure to organs at risk (OARs).

\subsection{MDP Formulation for RPT}

The proposed MDP framework encapsulates RPT dynamics with these components:

\begin{itemize}
    \item \textbf{State Space}: Constituted by Time-Integrated Activity (TIA) and absorbed doses in tumor and OARs, representing patient's physiological status and therapy impact
    \item \textbf{Action Space}: Corresponds to dosing strategies, including selection of radiopharmaceutical agents, dosages, and intervals
    \item \textbf{Reward Function}: Quantifies therapeutic efficacy and safety, maximizing tumor control while minimizing OAR radiation exposure
    \item \textbf{Transition Probabilities}: Employ Neural ODEs to model likelihood of moving from current state to new state given an action
\end{itemize}

\subsection{Decision Rule and Policy Development}

The decision rule within our MDP framework maps the current state to an optimal action, informed by cumulative knowledge in the Neural ODE model. We formalize policy $\pi$ as a sequence of decision rules across the treatment horizon.

\begin{algorithm}[!h]
\caption{Optimization of RPT Treatment via MDP and Neural ODEs}
\begin{algorithmic}[1]
\STATE Define state space based on TIA and absorbed doses
\STATE Establish action space for dosing strategies
\STATE Construct reward function prioritizing tumor control and OAR safety
\STATE Utilize Neural ODEs to determine transition probabilities
\STATE Initialize policy with baseline strategy
\WHILE{not converged}
    \STATE Evaluate current policy based on expected rewards
    \STATE Improve policy using policy iteration methods
\ENDWHILE
\STATE Deploy optimized policy for clinical decision support
\end{algorithmic}
\end{algorithm}

\subsection{Optimization Goal and Clinical Implementation}

The optimization goal is to calibrate policy $\pi$ to maximize expected utility of rewards across the patient's treatment trajectory, aligning with clinical objectives and patient-specific considerations.

In practical implementation, we plan to utilize a combination of model-free and model-based RL algorithms. Initial training will be performed through interaction with a Theranostic Virtual System (TVS) simulating real-world RPT dynamics. Algorithms like DQN and DDPG will be investigated alongside model-based techniques like Decision Transformers.

We propose offline training using historical patient data and simulated scenarios, which allows development and refinement without real-time experimentation. As the system matures and accumulates more real-world data, we plan to gradually transition to online learning, fine-tuning policies to adapt to new information.

\section{Performance Metrics and Evaluation Methodologies}

To ensure success of data-driven applications in theranostics, precise performance metrics, evaluation methodologies, and benchmarks are crucial\cite{neal2013response}\cite{dulac2021challenges}. Key metrics include:

\begin{itemize}
    \item \textbf{Sampling Efficiency}: Measures how much reward is lost before convergence to final performance
    \item \textbf{Delay in Observing}: Addresses challenges of delayed treatment effects, requiring methods to assign rewards that arrive significantly after causative events\cite{hung2019optimizing}
    \item \textbf{Continuous State and Action Spaces}: Requires frameworks like Action Elimination Deep Q Network (AE-DQN)\cite{zahavy2018learn} to handle high-dimensional spaces
    \item \textbf{Stochasticity}: Necessitates constrained MDPs framework for stochastic modeling with predefined environmental constraints\cite{altman1999constrained}
    \item \textbf{Explainability}: Essential for successful collaboration between algorithms and medical professionals
\end{itemize}

To measure RL algorithm reliability, methods have been proposed\cite{chan2019measuring} to evaluate reproducibility, stability, dispersion (width of distribution), and risk (heaviness of distribution's lower tail). Inter-Quartile Range (IQR) measures dispersion, while Conditional Value at Risk (CVaR) measures expected loss in worst-case scenarios.

Performance profiles\cite{agarwal2021deep} can be used for comparing algorithm performance, plotting score distribution across all runs and tasks with uncertainty estimates using stratified bootstrap confidence bands.

\section{Discussion and Future Work}

This study provided an overview of data-driven applications within RPT, identifying gaps and promising research avenues. Future endeavors should consider practicality of leveraging generic-driven models or system-of-systems frameworks as alternatives to abstracted ODE/PDE-based PBPK models.

Key questions for further exploration include:
\begin{enumerate}
    \item What is the most effective methodology to harmonize strengths of generic data-driven dynamics models with interpretability of ODE/PDE models?
    \item Does employing a comprehensive global system model confer advantages over a compartmentalized approach in accommodating inter-patient variability?
    \item Can principles of pre-trained large neural language models be transposed to model PBPK dynamics without diluting interpretability?
    \item How can RL be judiciously implemented in clinical settings, given its data-intensive nature and necessity for suitable representations?
    \item What strategies are most effective for evaluating models and decision-support tools while safeguarding patient well-being and conforming to ethical standards?
\end{enumerate}

As RPT continues to mature, these queries will guide the next wave of innovations. The convergence of computational approaches and biomedical expertise is set to unlock new therapeutic paradigms, tailoring treatment to individual patients' unique profiles. The promise of theranostics, empowered by data-driven approaches, heralds a new era of precision medicine where bespoke treatment becomes the standard of care.

\bibliographystyle{IEEEtran}

\end{document}